\documentclass[10pt,letterpaper]{article}

\usepackage[top=0.85in,left=1.0in,footskip=0.75in]{geometry}

\usepackage{lscape} 
\raggedright
\usepackage[english]{babel}
\usepackage{cite}
\usepackage{hyperref}

\usepackage{pdflscape}

\usepackage{amsmath,amssymb}

\usepackage{changepage}

\usepackage[utf8x]{inputenc}

\usepackage{textcomp,marvosym}

\usepackage{cite}

\usepackage{nameref,hyperref}

\usepackage[right]{lineno}

\usepackage{microtype}
\DisableLigatures[f]{encoding = *, family = * }

\usepackage[table]{xcolor}

\usepackage{array}

\newcolumntype{+}{!{\vrule width 2pt}}

\newlength\savedwidth

\usepackage[aboveskip=1pt,labelfont=bf,labelsep=period,justification=raggedright,singlelinecheck=off]{caption}

\makeatletter
\renewcommand{\@biblabel}[1]{\quad#1.}
\makeatother

\usepackage{tabularx}
\usepackage{ragged2e}
\newcolumntype{L}[1]{>{\raggedright\arraybackslash}p{#1}}
\newcolumntype{C}[1]{>{\centering\arraybackslash}p{#1}}
\newcolumntype{R}[1]{>{\raggedleft\arraybackslash}p{#1}}
\newcolumntype{J}[1]{>{\justifying\arraybackslash}p{#1}}
\usepackage{longtable}

\usepackage{lastpage,fancyhdr,graphicx}
\usepackage{epstopdf}
\fancyheadoffset[L]{2.25in}
\fancyfootoffset[L]{2.25in}
\usepackage{pst-tree,array} 
\usepackage{incgraph}

\usepackage{ragged2e}
\justifying

\begin{document}

\vspace*{0.2in}

\begin{flushleft}
{\Large
\textbf\newline{COMPUTER VISION FOR COVID-19 CONTROL: A SURVEY} 
}
\newline
\\
\textbf{Anwaar Ulhaq}\textsuperscript{1,2},
\textbf{Asim Khan}\textsuperscript{2},
\textbf{Douglas Gomes}\textsuperscript{2},
\textbf{Manoranjan Paul}\textsuperscript{1}
\\
\bigskip
\textbf{1} School of Computing and Mathematics, Charles Sturt University, NSW, Australia
\\
\textbf{2} College of Engineering and Science, Victoria University, VIC, Australia


\bigskip

\end{flushleft}


\section*{Abstract}
The  COVID-19 pandemic has triggered an urgent need to contribute to the fight against an immense threat to the human population. Computer Vision, as a subfield of Artificial Intelligence, has enjoyed recent success in solving various complex problems in health care and has the potential to contribute to the fight of controlling COVID-19. In response to this call, computer vision researchers are putting their knowledge base at work to devise effective ways to counter COVID-19 challenge and serve the global community. New contributions are being shared with every passing day. It motivated us to review the recent work, collect information about available research resources and an indication of future research directions. We want to make it available to computer vision researchers to save precious time. This survey paper is intended to provide a preliminary review of the available literature on the computer vision efforts against COVID-19 pandemic.


\section{Introduction}
COVID 19, an infectious disease is caused by severe acute respiratory syndrome (SARS-CoV-2)\cite{ref1} and named coronavirus due to visual appearance (under an electron microscope) to solar corona (similar to a crown)\cite{ref2}. The fight against COVID 19 has motivated researchers worldwide to explore, understand, and devise new diagnostic and treatment techniques to culminate this threat to our generation. In this article, we discuss how the computer vision community is fighting with this menace by proposing new types of approaches, improving efficiency, and speed of the existing efforts. \

Computer vision is an interdisciplinary field that deals with how computers can develop a high-level understanding by interpreting information present in digital images. It has made substantial progress in the last few years, mainly due to the success of deep learning, a sub-field of machine learning. Computer vision techniques have shown enormous scope in various applications areas, especially in healthcare and medical research. A plethora of novel computer vision approaches in healthcare applications exist that include, but not limited to, disease diagnosis, prognosis, surgery, therapy, medical image analysis and drug discovery\cite{ref5}. This success has enabled computer vision scientists to take the challenge as soldiers in the fight against COVID-19 by contributing to disease diagnosis, prognosis, prevention, control, treatment and management.\

The scientific response to combat COVID-19 has been far quicker and widespread. According to PubMed\cite{ref6}, in the year 2019, 755 academic papers published including the word “coronavirus” but even in the first 80 days of 2020 alone, the number rose to astounding 1,245 articles about the family of viruses. The economist has dubbed it “science of the times” with the hope that such efforts would help speed up the development of a COVOD-19 vaccine\cite{ref6}.   \

Numerous approaches in computer vision have been proposed so far, dealing with different aspects of combat the COVID-19 pandemic. These approaches differ from each other based on the way they approach the fundamental questions: How can medical imaging facilitate faster and reliable diagnosis of COVID-19? Which image features properly classify conditions as Bacterial, Viral, COVID-19, and Pneumonia? What can we learn from imaging data acquired from disease survivors to screen critical and non-critical ill patients? How can computer vision be used to enforce social distancing and early screening of infected people? How can 3D computer vision help to maintain healthcare equipment supply and guide the development of a COVID-19 vaccine? The answers to these questions are being explored, and preliminary work has been done.  The goal of this survey is to group computer vision methods into broad categories and provide comprehensive descriptions of representative methods in each category. We aspire to give readers the ability to understand the baseline efforts and kickstart their work where others have left.  Moreover, we aim to identify new trends and ideas to build a more robust and well-planned strategy in this war of our times.\\

\begin{figure}[!htb]
        \center{\includegraphics[width=\textwidth]
        {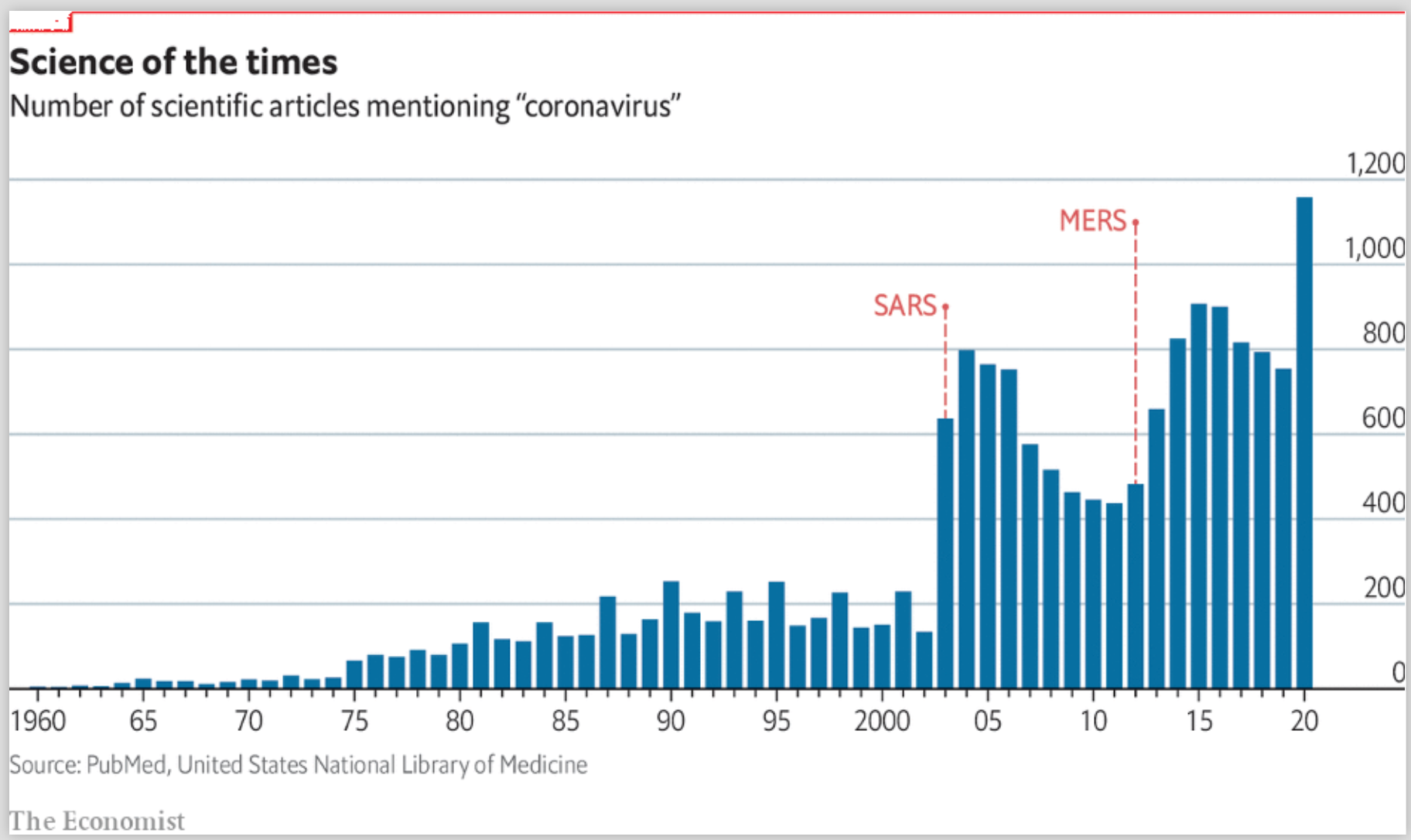}}
        \caption{\label{fig:my-label} A portrayal of current increase in research articles about coronavirus related research. Adapted from\cite{ref6}. }
      \end{figure}

Our survey will include research papers in pre-print format due to time urgency imposed by this disease. It is not an optimal approach due to the risk of lower quality and work without due validation. Many of the works have not been put into clinical trial as it is time-consuming. Nevertheless, our intention here is to share ideas from a single platform while highlighting the computer vision community efforts. We hope that our reader is aware of these contemporary challenges.  \

We follow a bottom-up approach to describing the research problems that need addressing first. We start with disease diagnosis, discuss disease prevention and control, followed by treatment-related computer vision research work. Section 2 describes the overall taxonomy of computer vision research areas by classifying these efforts into three classes. Section 3 provides a detailed description of each research area, relevant papers and a brief description of representative work. Section 4 describes available resources including research datasets, their links, deep learning models and codes. Section 5 provides the discussion and future work directions followed by concluding remarks and references. \


\section{HISTORICAL DEVELOPMENT }

The novel coronavirus SARS-CoV-2 (previously known as 2019-nCoV ) is the seventh member of the Coronaviridae family of viruses which are enveloped, non-segmented, positive-sense RNA viruses\cite{ref3}. The mortality rate of COVID-19 is less than that of the severe acute respiratory syndrome (SARS) and Middle East respiratory syndrome (MERS) coronavirus diseases (10\% for SARS-CoV and 37\% for MERS-CoV). However, it is highly infectious, and the number of cases has been increasing rapidly\cite{ref10}. \

The disease outbreak first reported in Wuhan, the Hubei province of China\cite{ref3}, after several cases of pneumonia with unknown causes were reported on 31 December 2019. A novel coronavirus was discovered as the causative organism through deep sequencing analysis of samples of patients’ respiratory tract at Chinese facilities on 7 January 2020\cite{ref10}. The outbreak was declared a Public Health Emergency of International Concern on 30 January 2020. On 11 February 2020, the World health organization (WHO)announced a name for the new coronavirus disease: COVID-19. It was officially being considered pandemic after 11 March announcement by WHO\cite{ref4}.  \


\section{TAXONOMY OF KEY AREAS OF RESEARCH}

In this section, we describe the classification of computer vision techniques that try to counter the menace of COVID-19. For better comprehensibility, we have classified them into three key areas of research: (i) diagnosis and prognosis, (ii) disease prevention and control, and (iii) disease treatment and management. This taxonomy is shown in Figure 2. In the following subsections, we discuss the research fields, the relevant papers, and present a brief representative description of related works. \

\begin{figure}[!htb]
        \center{\includegraphics[width=\textwidth]
        {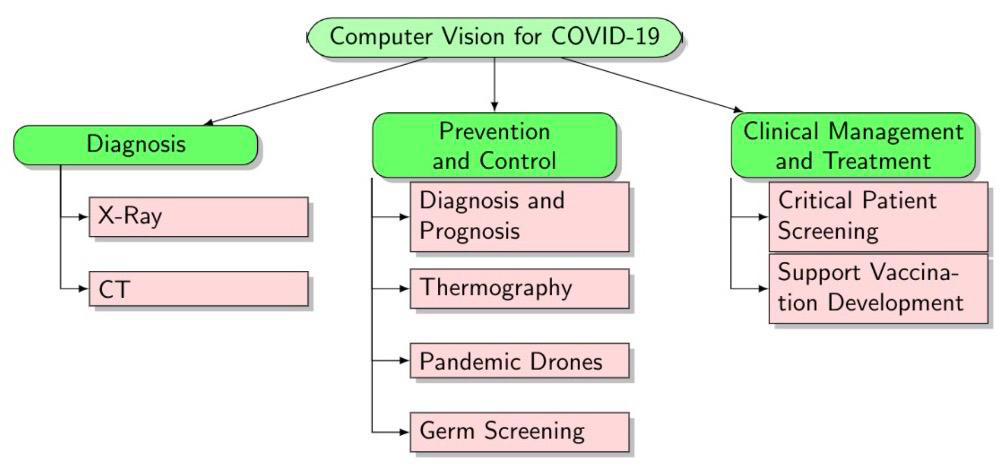}}
        \caption{\label{fig:my-label} A classification Computer Vision Approaches for COVUD-19 Control }
      \end{figure}



\subsection{Diagnosis and Prognosis}
An essential step in this fight is the reliable, faster and affordable diagnostic process that can be readily accessible and available to the global community. According to Cambridge dictionary\cite{ref15}, diagnosis is "the making of a judgment about the exact character of a disease or other problem, esp. after an examination, or such a judgment" and prognosis is "a doctor's judgment of the likely or expected development of a disease or of the chances of getting better".\

Currently, Reverse transcriptase quantitative polymerase chain reaction (RT-qPCR) tests are considered as the gold standard for diagnosing COVID-19\cite{ref7}. During such a test, small amounts of viral RNA are extracted from a nasal swab, amplified, quantified. Virus detection is then performed using a fluorescent dye. Although accurate, the test is time-consuming and manual, which limits its availability in large scales. Some studies have also shown false-positive PCR testing\cite{ref8}.\


\subsubsection{Computed Tomography (CT) scan}

An alternative approach is the use of radiology examination that uses computed tomography (CT) imaging\cite{ref9}. A chest CT scan is a non-invasive test conducted to obtain a precise image of a patient’s chest. It uses an enhanced form of x-ray technology, providing more detailed images of the chest than a standard x-ray. It produces images that include bones, fats, muscles, and organs, giving physicians a better view, which is crucial when making accurate diagnoses. \

There are two types of chest CT scan, namely the high-resolution and spiral chest CT scan\cite{ref16}. The high-resolution chest CT scan provides more than a slice (or image) in a single rotation of the x-ray tube. The spiral chest CT scan application involves a table that continuously moves through a tunnel-like hole while the x-ray tube follows a spiral path. The advantage of the spiral CT is that it is capable of producing a three-dimensional image of the lungs. \


\begin{figure}[!htb]
        \center{\includegraphics[width=\textwidth]
        {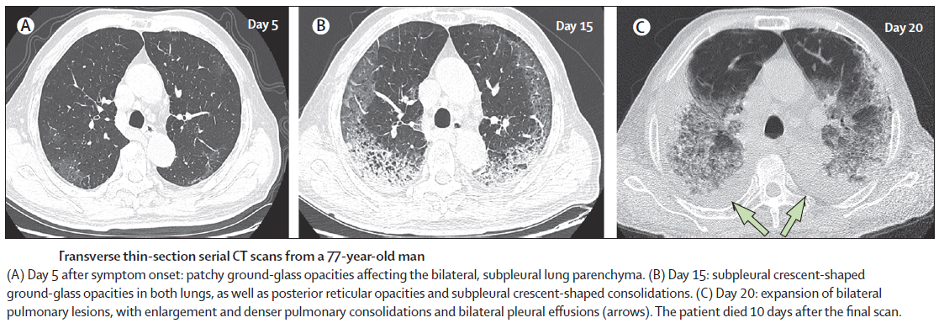}}
        \center{\includegraphics[width=\textwidth]
        {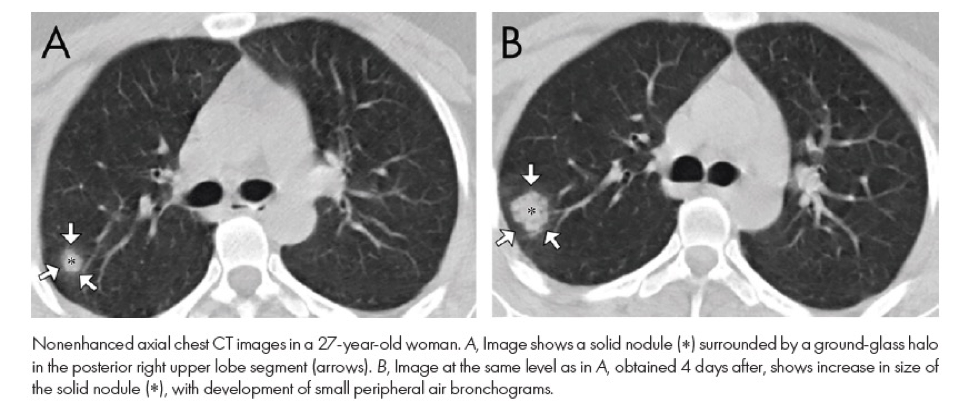}}
        \caption{\label{fig:my-label} CT images adapted from\cite{ref10} \cite{ref20} portray CT features related to COVID-19. Ground glass opacities (top) and ground glass halo (bottom). }
      \end{figure}

Important CT features include ground-glass opacity, consolidation, reticulation/thickened interlobular septa, nodules, and lesion distribution (left, right or bilateral lungs)\cite{ref17} \cite{ref18} \cite{ref19} \cite{ref36}. The most observable CT features discovered in COVID-19 pneumonia are bilateral and subpleural areas of ground-glass opacification, consolidation affecting the lower lobes. Within the intermediate stage (4–14 days from symptom onset), crazy-paving pattern and possibly observable Halo sign become important features as well\cite{ref8} \cite{ref9} \cite{ref10} \cite{ref9} \cite{ref17} \cite{ref18} \cite{ref19} \cite{ref20} \cite{ref36}. One case of CT images is shown in figure 3. As the identification of disease features is time-consuming, even for expert radiologists, computer vision can help by automating such a process.\


\subsubsection*{Representative Work, Evaluation and Discussion}

To date, various CT-scanning automated approaches have been proposed\cite{ref9} \cite{ref12} \cite{ref13} \cite{ref14} \cite{ref15} \cite{ref16} \cite{ref17} \cite{ref18} \cite{ref19} \cite{ref20} \cite{ref21} \cite{ref22} \cite{ref23} \cite{ref24} \cite{ref25}. To discuss the approach and performance of the computer vision CT-based disease diagnosis, we have selected some recent representative works that provide an overview of their effectiveness. It is worth noting that they have been presenting different performance metrics and using a diverse number of images and datasets. These practices make their comparison very challenging. Some of the metrics include Accuracy, Sensitivity, Specificity, Area Under Curve (AUC), Positive predictive value (PPV), Negative predictive value (NPV), and F1 score. A quick elucidation on their definition can be useful. The accuracy of a method determines how correct the values are predicted. The precision determines the reproducibility of the measurement or how many of the predictions are correct. Recall shows how many of the correct results are discovered. F1-score uses a combination of precision and recall to calculate a balanced average result. 

The first class of work discussed here approaches diagnosis as a segmentation problem. Jun Chen et al.\cite{ref21} has proposed a CT image dataset of 46,096  images of both healthy and infected patients, labeled by expert radiologists. It was collected from 106 patients admitted with 51 confirmed COVID-19 pneumonia and 55 control patients. The work used deep learning models for segmentation only so that it could identify the infected area in CT images between healthy and infected patients. It was based on UNet++ semantic segmentation model\cite{ref22}, used to extract valid areas in the images. It used 289 randomly selected CT images and tested it on other 600 randomly selected CT images. The model achieved a per-patient sensitivity of 100\%, specificity of 93.55\%, accuracy of 95.24\%, PPV (positive prediction value) of 84.62\%, and NPV(negative prediction value) of 100\%. In the retrospective dataset, it resulted in a per-image sensitivity of 94.34\%, specificity of 99.16\%, accuracy of 98.85\%, PPV of 88.37\%, and NPV of 99.61\%.  The trained model from this study was deployed at the Renmin Hospital of Wuhan University (Wuhan, Hubei province, China) to accelerate the diagnosis of new COVID-19 cases. It was also open-sourced on the Internet as to enable rapid review of new cases in other locations. A cloud-based open-access artificial intelligence platform was constructed to provide support for detecting COVID-19 pneumonia worldwide. For this purpose, a website has been made available to provide free access to the present model at (\url{http://121.40.75.149/znyx-ncov/index}). \

The second type of work considered COVID-19 as a binary classification problem. Lin Li\cite{ref23} proposed (COVNet), to extract visual features from volumetric chest CT using transfer learning on the RESNET50. Lung segmentation was performed as a pre-processing task using the U-Net model. It used 4356 chest CT exams from 3,322 patients from the dataset collected from 6 hospitals between August 2016 and February 2020. The sensitivity and specificity for COVID-19 are 90\% (114 of 127; p-value<0.001) with 95\% confidence interval (CI) of [95\% CI: 83\%, 94\%] and 96\% (294 of 307; p-value<0.001) with [95\% CI: 93\%, 98\%], respectively. The model was also made available online for public use at \url{https://github.com/bkong999/COVNet}. \

The diagnosis problem was also approached as a 3-category classification task: distinguishing healthy patients from those with other types of pneumonia and those with COVID-19. Song et al.\cite{ref23} used data from 88 patients diagnosed with the COVID-19, 101 patients infected with bacteria pneumonia, and 86 healthy individuals. It proposed DRE-Net (Relation Extraction neural network) based on ResNet50, on which the Feature Pyramid Network (FPN)\cite{ref24} and the Attention module were integrated to represent more fine-grained aspects of the images. An online server is available for online diagnoses with CT images at \url {http://biomed.nsccgz.cn/server/Ncov2019.} \

Due to limited time available for annotations and labelling, weakly-supervised deep learning-based approaches have also been developed using 3D CT volumes to detect COVID-19. Chuansheng Zheng\cite{ref25} proposed 3D deep convolutional neural Network (DeCoVNet) to Detect COVID-19 from CT volumes. The weakly supervised deep learning model could accurately predict the COVID-19 infectious probability in chest CT volumes without the need for annotating the lesions for training. The CT images were segmented using a pre-trained UNet. It used 499 CT volumes for training, collected from 13 December 2019 to  23 January 2020, and 131 CT volumes for testing, collected from 24 January 2020 to 6 February 2020. The authors chose a probability threshold of 0.5 to classify COVID-positive and COVID-negative cases. The algorithm obtained an accuracy of 0.901, a positive predictive value of 0.840, and a high negative predictive value of 0.982. The developed deep learning model is available at \url{https://github.com/sydney0zq/covid-19-detection}.



\begin{longtable}{|L{1.5cm}|L{2.5cm}|L{2.5cm}|L{2cm}|L{2.5cm}|L{2cm}|L{2.5cm}|} \hline

\caption {\bf : Representative works for CT based COVID-19 Diagnosis}\\

\hline

 \bf Study&  \bf Classification Model and availability & \bf Segmentation Model & \bf Dataset & \bf No of Participants & \bf Deployed & \bf Performance \\ \hline 
 
\endfirsthead
\hline
\multicolumn{4}{c}%
{\tablename\ \thetable\ -- \textit{Continued from previous page}} \\
\hline
\bf Study&  \bf Classification Model and availability & \bf Segmentation Model & \bf Dataset & \bf No of Participants & \bf Deployed & \bf Performance \\ \hline
\endhead
\hline \multicolumn{4}{r}{\textit{Continued on next page}} \\
\endfoot
\hline
\endlastfoot

Jun Chen et.al.\cite{ref21} & \url {(http://121.40.75.149/znyx-ncov/index) } & UNet++ to extract valid areas in CT images using 289 randomly selected CT images  & 46,096  CT images  & 106 patients with 51 confirmed COVID-19 pneumonia  & Renmin Hospital of Wuhan University (Wuhan, Hubei province,
China) & sensitivity of 100\%, specificity of 93.55\%, accuracy of 95.24\%  \\ 
\hline 

Shuai Wang et al.\cite{ref26} &  Modified inception\cite{ref27} with transfer learning available at : \url {https://ai.nscc-tj.cn/thai/deploy/public/pneumonia_ct.}
  & - & 453 CT images of pathogen-confirmed COVID-19 & 99 patients from Xi’an Jiaotong University First Affiliated
Hospital, Nanchang University First Hospital and Xi’An No.8 Hospital of Xi’An
Medical College & - & The internal validation achieved accuracy of 82.9\% with specificity of 80.5\% and sensitivity of 84\%. 
The external testing dataset showed  accuracy of 73.1\% with specificity of 67\%.
 \\ \hline 

Xiaowei Xu et al.\cite{ref28} &   Combination of Two CNN three-dimensional classification models (ResNet-18 network\cite{ref29} +location-attention oriented model). & VNET based segmentation model\cite{ref30} &  A total of 618 CT samples were collected: 19, 224 CT samples Flue l175 CT samples from healthy people & 219 from 110 patients with COVID- and 224 patients with Influenza-A viral pneumonia & - & Model accuracy  86.7\%  \\ 
\hline

Ying Song et al.\cite{ref31} &  Details Relation Extraction neural networkDRE-Net + ResNet50\cite{ref32}, with Feature Pyramid Network (FPN)+ Attention module.
An online server is available for online diagnoses with CT images by \url{ http://biomed.nsccgz.cn/server/Ncov2019.} & - & 777 CT images & 88 patients diagnosed with the COVID-19101 patients infected with bacteria pneumonia, and 86 healthy persons & - & AUC of 0.99 and recall (sensitivity) of 0.93. Accuracy of 0.86 and  F-Score 0.87
 \\ \hline

Ophir Gozes et al.\cite{ref33} &  2D deep convolutional neural network architecture based on Resnet-50 \cite{ref32} & U-net architecture for image segmentation\cite{ref34} & - & 56 patients with confirmed COVID-19 diagnosis & - & 0.996 AUC (95\%CI: 0.989-1.00) on Chinese control and infected patients. 98.2\% sensitivity, 92.2\% specificity. \\ 
\hline

Fei Shan\cite{ref40} &  -  & VB-Net” neural network to segment COVID-19 infection regions in CT scans & 249 CT images & 249 COVID-19 patients, and validated using new COVID-19 patients & - & Dice similarity coefficients of 91.6\% ± 10.0\% between automatic and manual segmentations \\ 
\hline

Cheng Jin et al\cite{ref36} &  Model is available at : \url {https : =github:com=ChenWWWeixiang=diagnosis_covid19}.  & 2D CNN  based AI system, model name is not specified & 970 CT volumes & 496 patients with confirmed COVID-19  & - & accuracy of 94.98\%, an area under the receiver operating characteristic curve (AUC) of 97.91\% \\ 
\hline

 Mucahid Barstugan1 et.al.\cite{ref38} &  Grey Level Co-occurrence Matrix Local Directional Pattern Grey Level Run Length Matrix Grey Level Size Zone Matrix 
Discrete Wavelet Transform + SVM & - &  150 CT images. & - & - & 99.68\% classification accuracy \\ 
\hline

Lin Li\cite{ref23} &  (COVNet), was developed to extract visual features from volumetric chest CT RESNET50. Model is available at : \url {https://github.com/bkong999/COVNet} & U-Net for segmentation & 4356 chest CT images & The datasets were collected from 6 hospitals and 3,322 patients. & - & The sensitivity and specificity for COVID-19 are 90\%  and 96\% respectively. \\ 
\hline

Chuansheng Zheng\cite{ref41} &  3D deep convolutional neural Network to Detect COVID-19 (DeCoVNet) from CT volumes. The developed deep learning software is available at \url {https://github.com/sydney0zq/covid-19-detection.} & Segmented using a pre-trained UNet & - & Union Hospital, Tongji Medical College, Huazhong University of Science and Technology). Finally, 540 patients & - & Obtained 0.959 ROC AUC and 0.976 PR AUC. \\
 \hline

Shuo Jin\cite{ref35} &  Transfer learning on ResNet-50  & segmentation model as 3D
U-Net++,  & - & Using 1,136 training cases (723 positives for COVID-19) from five hospitals & Deployed the system in 16 hospitals in China & AUC  0.991 sensitivity of 0.974 and specificity of 0.922.
 \\ \hline

\end{longtable}



\subsubsection{X-ray Imagery}  

The drawback of using CT imaging is the need for high patient dose and enhanced cost\cite{ref42}. It makes digital chest x-ray radiography (CXR) as the imaging modality with the lower cost and wider availability for detecting chest pathology. Therefore, automated diagnosis of COVID-19 features in CXR will make it a highly useful diagnostic tool against the disease.
Digital X-ray imagery computer-aided diagnosis is used for different diseases, including osteoporosis\cite{ref43}, cancer\cite{ref44} and cardiac disease\cite{ref45}.  However, as it is really hard to distinguish soft tissue with a poor contrast in X-ray imagery, contrast enhancement is used as pre-processing step\cite{ref46} \cite{ref47}. Lung segmentation of chest X-rays is a crucial and important step in order to identify lung nodules and various segmentation approaches are proposed in the literature\cite{ref48} \cite{ref49} \cite{ref50} \cite{ref51}. \

CXR examinations have shown consolidation in COVID-19 infected patients. In one study at Hong Kong\cite{ref52}, three different patients had daily CXR, two of them showed progression in the lung consolidation over 3-4 days. Further CXR examinations show improvement over the subsequent two days. The third patient showed no significant changes over an 8-day period. However, a similar study showed that the ground glass opacities in the right lower lobe periphery on the CT are not visible on the chest radiograph, which was taken 1 hour apart from the first study. However, CXR is still recommended alongside CT for better radiological analysis. Various CXR-related automated approaches have been proposed. The following section discusses the most salient work while Table 2 presents a more systematic presentation of such methods.\

\subsubsection*{Representative Work, Evaluation and Discussion}

To date, many deep learning-based computer vision models for X-ray COVID-19 were proposed. One of the most significant development is the model COVID-Net\cite{ref54} proposed by Darwin AI, Canada. In this work, human-driven principled network design prototyping is combined with machine-driven design exploration to produce a network architecture for the detection of COVID-19 cases from chest X-ray. The first stage of the human-machine collaborative design strategy is based on residual architecture design principles. The dataset used to train and evaluate COVID-Net, is referred to as COVIDx\cite{ref54} and comprise a total of 16,756 chest radiography images across 13,645 patient cases. The proposed model achieved 92.4\% accuracy 80\% sensitivity for COVID-19 diagnosis. \

\begin{figure}[!t]
        \center{\includegraphics[width=\textwidth]
        {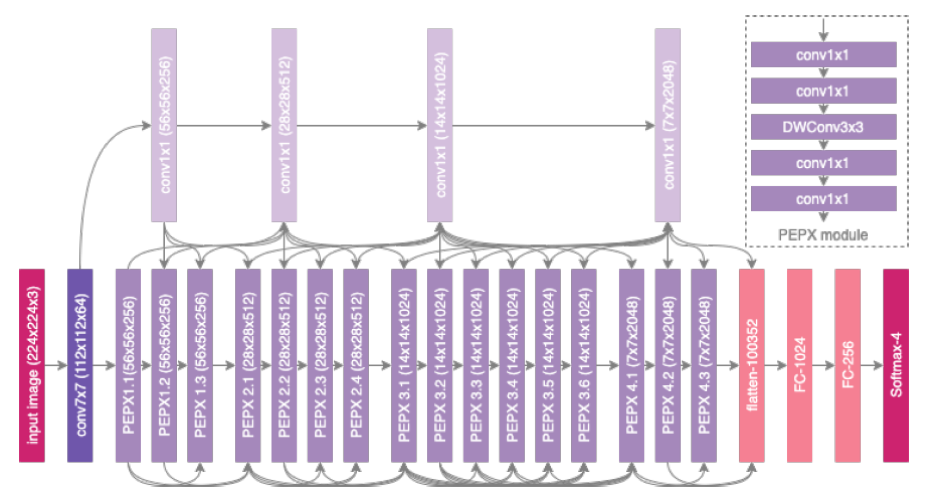}}
        \caption{\label{fig:my-label} Architectural diagram of COVID-Net\cite{ref54}. High architectural diversity and selective long-range connectivity can be observed. }
        \label{archdiag}
      \end{figure}

The initial network design prototype makes one of three classes: a) no infection (normal), b) non-COVID19 infection (viral and bacterial), and c) COVID-19 viral infection. The goal is to aid clinicians to better decide which treatment strategy to employ depending on the cause of infection, since COVID-19 and non-COVID19 infections require different treatment plans. In the second stage, data, along with human-specific design requirements, act as a guide to a design exploration strategy to learn and identify the optimal macro- and microarchitecture designs to construct the final tailor-made deep neural network architecture. The proposed COVIDNet network diagram is shown in Fig. \ref{archdiag} and available publicly at \url{https://github.com/lindawangg/COVID-Net}.\


Ezz El-Din Hemdan et al.\cite{ref55} proposed the COVIDX-Net based on seven different architectures of DCNNs; namely VGG19, DenseNet201\cite{ref56}, InceptionV3, ResNetV2, InceptionResNetV2, Xception, and MobileNetV2\cite{ref57}. These models were trained on COVID-19 cases provided by Dr Joseph Cohen and Dr Adrian Rosebrock, available at \url {https://github.com/ieee8023/covid-chestxray-dataset}\cite{ref58}. Tthe best model combination resulted in F1-scores of 0.89 and 0.91 for normal and COVID-19 cases. Similarly, Asmaa Abbas et al.\cite{ref59} proposed a Decompose, Transfer, and Compose (DeTraC) approach for the classification of COVID-19 chest X-ray images. The authors applied CNN features of pre-trained models on ImageNet and ResNet to perform the diagnoses. The dataset consisted of 80 samples of normal CXRs (with 4020 x 4892 pixels) from the Japanese Society of Radiological Technology (JSRT) Cohen JP. COVID-19 image data collection, available at \url {https://githubcom/ieee8023/covid-chestxray-dataset}\cite{ref58}. This model achieved an accuracy of 95.12\% (with a sensitivity of 97.91\%, a specificity of 91.87\%, and a precision of 93.36\%). The code is available at \url{https://github.com/asmaa4may/DeTraC COVId19}.\

Uncertainty-Aware COVID-19 Classification and Referral model was introduced by tBiraja Ghoshal et al.\cite{ref60} with the proposed Dropweights based on Bayesian Convolutional Neural Networks (BCNN). In order for COVID-19 detection to be meaningful, two tyes of predictive uncertainty in deep learning were used on a subsequent work\cite{ref61}. One of it is Epistemic or Model uncertainty accounts for the model parameters uncertainty as it does not take all of the aspects of the data into account or the lack of training data. The other is Aleatoric uncertainty that accounts for noise inherent in the observations due to class overlap, label noise, homoscedastic and heteroscedastic noise, which cannot be reduced even if more data were to be collected. Bayesian Active Learning by Disagreement (BALD)\cite{ref62}, is based on mutual information that maximises the information between model posterior and predictions density functions approximated as the difference between the entropy of the predictive distribution and the mean entropy of predictions across samples.\


\begin{figure}[!t]
        \center{\includegraphics[width=\textwidth]
        {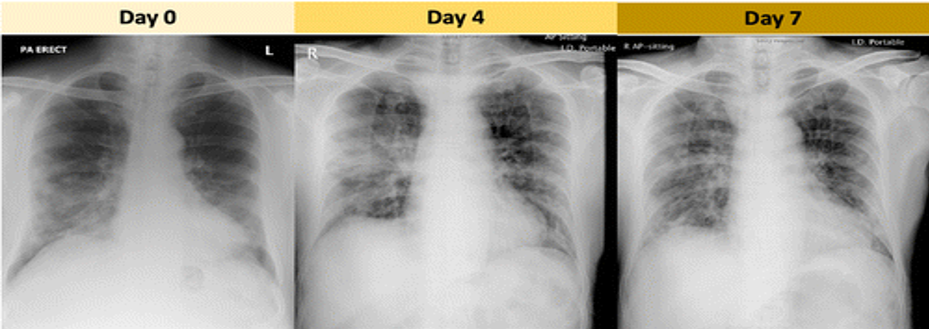}}
        \caption{\label{fig:my-label} Chest radiographs of an elderly male patient from Wuhan, China, who travelled to Hong Kong, China. These are 3 chest radiographs selected out of the daily chest radiographs acquired in this patient. The consolidation in the right lower zone on day 0 persist into day 4 with new consolidative changes in the right midzone periphery and perihilar region. This midzone change improves on the day 7 film. Image adapted from\cite{ref52}. }
      \end{figure}

A BCCN model was trained on 68 Posterior-Anterior (PA) X-ray images of lungs with COVID-19 cases from Dr Joseph Cohen’s Github repository\cite{ref58}, augmented the dataset with Kaggle’s Chest X-Ray Images (Pneumonia) from healthy patients. It achieved 88.39\% accuracy on the available dataset. This work additionally recommended visualisation of distinct features, as an additional insight to point prediction for a more informed decision-making process. It used the saliency maps produced by various state-of-the-art methods, e.g. Class Activation Map (CAM)\cite{ref63}, Guided Backpropagation and Guided Gradient, and Gradients to show more distinct features in the CSR images.\


\begin{figure}[!htb]
        \center{\includegraphics[width=\textwidth]
        {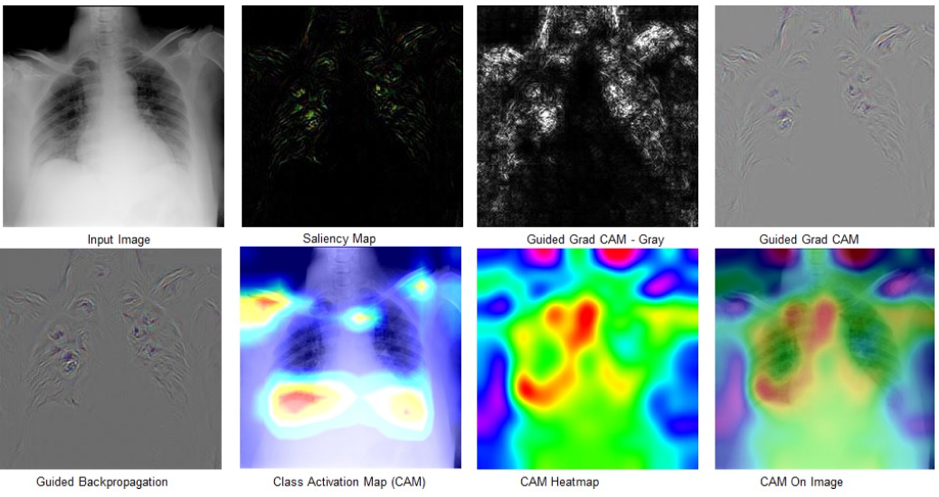}}
        \caption{\label{fig:my-label}Visualizations shown by using different saliency maps that provide additional insights diagnosis. Adapted from\cite{ref60}.}
      \end{figure}




\begin{longtable}{|L{3cm}|L{4cm}|L{4cm}|L{4cm}|}
\hline
\caption {\bf : Representative work for X-ray based COVID-19 Diagnosis}\\
\hline
 \bf Study&  \bf Model & \bf Dataset & \bf Performance \\ \hline 

\endfirsthead
\hline
\multicolumn{4}{c}%
{\tablename\ \thetable\ -- \textit{Continued from previous page}} \\
\hline
\bf Study&  \bf Model & \bf Dataset & \bf Performance \\ \hline
\endhead
\hline \multicolumn{4}{r}{\textit{Continued on next page}} \\
\endfoot
\hline
\endlastfoot

Guszt´av Ga´al et al.\cite{ref51} & Attention U-Net+ adversarial+ Contrast Limited Adaptive Histogram Equalization (CLAHE)\cite{ref65} & 247 images from Japanese Society of Radiological Technology (JSRT) Dataset+ Shenzhen dataset contains a total of 662 chest X-rays & DSC of 97.5\% on the JSRT dataset \\ \hline 

Asmaa Abbas et al.\cite{ref59} &  CNN features of pre-trained models on ImageNetand ResNet+ Decompose, Transfer, and Compose (DeTraC), for the
classification of COVID-19 chest X-ray images: The developed code is available at \url {https://github.com/asmaa4may/DeTraC COVId19} & I80 samples of normal CXRs (with 4020 x 4892 pixels) from the Japanese Society of Radiological Technology (JSRT) + Cohen JP. COVID-19 image data collection. \url {https://githubcom/ieee8023/covid-chestxray-dataset. 2020;}. &  High accuracy of 95.12\% (with a sensitivity of 97.91\%, a specificity of 91.87\%, and a precision of 93.36\%) \\ 
\hline

Ali Narin et al.\cite{ref66} &  Pre-trained ResNet50 model with transfer learning & The open source GitHub repository shared by Dr. Joseph Cohen+Chest X-Ray Images (Pneumonia) \url {https://www.kaggle.com/paultimothymooney/chest-xray-pneumonia}  &  accuracy
 (97\% accuracy for InceptionV3 and 87\% accuracy for Inception-ResNetV2).
 \\ \hline

Linda Wang et al. \cite{ref53} &  COVID-Net: lightweight residual projection expansion-
projection-extension (PEPX) design pattern, Model is 
available publicly for open access at \url {https://github.com/lindawangg/COVID-Net.}
 & COVIDx dataset: 16,756 chest radiography images across 13,645 patient cases from two open access data repositories &  Accuracy 92.4\%on COVIDx dataset \\ \hline

 Ezz El-Din Hemdan et al.\cite{ref55} &  COVIDX-Net: based on seven different architectures of DCNNs; namely VGG19, DenseNet201, InceptionV3, ResNetV2, Inception ResNetV2, Xception, and MobileNetV2  & COVID-19 cases provided by Dr. Joseph Cohen and Dr. Adrian Rosebrock\cite{ref58} &  F1-scores of 0.89 and 0.91 for normal and COVID-19, respectively \\ \hline

 Khalid EL ASNAOUI et al.\cite{ref69} &  
 Fined tuned versions of (VGG16, VGG19, DenseNet201, Inception-ResNet-V2, Inception-V3, Resnet50, MobileNet-V2 and Xception). & 5856 images (4273 pneumonia and 1583 normal).&  Resnet50, MobileNet-V2 and Inception-Resnet-V2 show highly satisfactory performance with accuracy (more than 96\%).
 \\ \hline 

Prabira Kumar Sethy et al \cite{ref68} &  Deepfeatures fromResnet50 + SVM classification & Data available in the repository of GitHub, Kaggle and Open-i as per their validated X-ray images. &  resnet50 plus
SVM achieved accuracy, FPR, F1 score, MCC and Kappa are 95.38\%,95.52\%, 91.41\% and 90.76\%
Respectively.
 \\ \hline

Ioannis D. Apostolopoulos1 et al.\cite{ref67} &  Various fine-tune dmodels: VGG19, MobileNet, Inception,Inception Resnet V2, Xception & 1427 X-Ray images. 224 images with confirmed COVID-19, 700 images with confirmed common pneumonia, and 504 images of normal conditions are included &  Accuracy with Xception was the highest, 95.57, sensitivity of 0.08 and specificity of 99.99.
 \\ \hline

Biraja Ghoshal et al.\cite{ref60} &  Dropweights based Bayesian Convolutional Neural
Networks (BCNN)
 & 68 Posterior-Anterior (PA) X-ray images of lungs with COVID-
19 cases from Dr. Joseph Cohen’s Github repository, augmented the dataset with Kaggle’s Chest X-Ray Images (Pneumonia) from healthy patients, a total of 5941 PA chest radiography images across 4 classes (Normal: 1583, Bacterial Pneumonia: 2786, non-COVID-19 Viral Pneumonia: 1504, and COVID-19: 68).
 &  88.39\% accuracy with BCNN \\ \hline

Muhammad Farooq, Abdul Hafeez\cite{ref54} &  3-step technique to fine-tune a pre-trained ResNet-50 architecture to improve model performance & COVIDx dataset  &   Accuracy of 96.23\% (on all the classes) on the COVIDx dataset  \\ \hline


\end{longtable}



\subsection{Prevention and Control} \
WHO has provided some guidelines on infection prevention and control (IPC) strategies for use when infection with a novel coronavirus is suspected\cite{ref70}. Major IPC strategies to limit transmission in health care settings include early recognition and source control, applying standard precautions for all patients; implementing additional empiric precautions like airborne precautions for suspected cases of COVID-19; implementing administrative controls and using environmental and engineering controls. Computer vision applications are providing excellent support for the implementation of IPC strategies. \

\subsubsection*{Representative Work, Evaluation and Discussion}
\
The use of masks or protective equipment to limit the virus spread was a strategy identified in the early stage of disease progression. Some countries, China being the most prominent example, implemented it as a control strategy. Computer vision-based systems greatly facilitated such implementation. \

Zhongyuan Wang et al.\cite{ref71} proposed Masked Face Recognition approach using a multigranularity masked face recognition model, resulting in 95\% accuracy on a masked face images dataset. The data was made public for research and provide three types of masked face datasets, including Masked Face Detection Dataset (MFDD)\cite{ref72}, Real-world Masked Face Recognition Dataset (RMFRD) and Simulated Masked Face Recognition Dataset (SMFRD)\cite{ref73}. \


\begin{figure}[!htb]
        \center{\includegraphics[width=\textwidth]
        {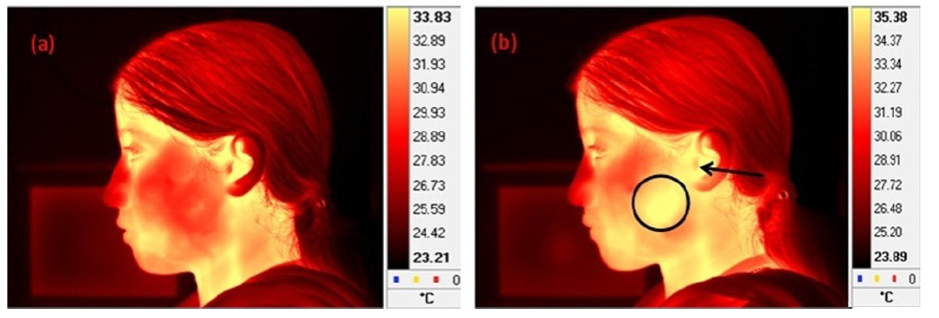}}
        \caption{\label{fig:my-label} Temperature screening in process using thermal images of a subject talking on a hand-held mobile phone; (a) after 1 min and (b) after 15 min of talking. After 15 min of talking the temperature of the encircled region increased from 30.56 to 35.15 C, whereas the temperature of the ear region (indicated by an arrow) increased from 33.35 to 34.82C. Similar system can be used for fever screening.}
      \end{figure}

Infrared thermography was also recommended as an early detection strategy for infected people, especially in crowns like passengers on an airport. A comprehensive review of medical applications of infrared thermography is provided by B.B. Lahiri [74] \cite{ref74}, including fever screening. Somboonkaew et al.\cite{ref75} proposed a mobile-platform for an automatic fever screening system based on infrared forehead temperature. Best practices for standardized performance and testing of infrared thermographs intended for fever screening are discussed by Ghassemi et al. \cite{ref76}. Negishi T \cite{ref77} proposed an Infection Screening System Using Thermography and CCD Camera with Good Stability and Swiftness for Non-contact Vital-Signs Measurement by Feature Matching and MUSIC Algorithm.  Earlier for SARD spread control, Chiu et al.\cite{ref78} proposed a computer vision systems help in fever screening, which was used in earlier outbreaks of SARS.  From 13 April to 12 May 2003, 72,327 patients and visitors passed through the only entrance allowed at TMU-WFH where a thermography station was in operation. \

Additional miscellaneous approaches for prevention and control are also worth noting. An example is pandemic drones using remote sensing and digital imagery, which were recommended for identifying infected people. Al-Naji et al.\cite{ref79} have used such a system for remote life sign monitoring in disaster management in the past. A similar application is to use vision-guided robot control for 3D object recognition and manipulation. Moreover, 3D modelling and printers are helping to maintain the supply of healthcare equipment in this troubled time. Joshua M. Pearce\cite{ref81} discusses RepRap-class 3-D printers and open-source microcontrollers. The applications are relevant since mass distributed manufacturing of ventilators has the potential to overcome medical supply shortages. Lastly, germ scanning is an important step against combating COVID-19. Edouard A. Hay\cite{ref82} have proposed a convolutional neural network for germ scanning such as the identification of bacteria Light-sheet microscopy image data with more than 90\% accuracy. \


\begin{longtable}{|L{2cm}|L{5cm}|L{6cm}|L{2.2cm}|} \hline
\caption {\bf : Representative works for infected disease Prevention and Control} \\
\hline
 \bf Study&  \bf Prevention Control Methodology& \bf Implementation/ Dataset& \bf Performance \\ \hline
 
 \endfirsthead
\hline
\multicolumn{4}{c}%
{\tablename\ \thetable\ -- \textit{Continued from previous page}} \\
\hline
\bf Study&  \bf Prevention Control Methodology & \bf Implementation/ Datase & \bf Performance \\ \hline
\endhead
\hline \multicolumn{4}{r}{\textit{Continued on next page}} \\
\endfoot
\hline
\endlastfoot
  
Zhongyuan Wang et al.\cite{ref71} &Masked Face Recognition based on deep learning  & Dataset is available at:\ \url {https://github.com/X-zhangyang/ Real-World-Masked-Face-Dataset.} &The multigranularity masked face recognition model we developed achieves 95\% accuracy  \\ 
\hline 

 Joshua M. Pearce\cite{ref81} & RepRap-class 3-D printers and open source microcontrollers, mass distributed  manufacturing of ventilators &  &  3D printing can facilitate supply chain.\\
  \hline 
 
 W. Chiu, et al.\cite{ref78} & Inrfrared thermograpgy : Mass-fever screening &72,327 patients or visitors passed through the only entrance where a thermography station was in operation.&Over a period of one month, hundred and five patients or visitors were detected to have a thermographic fever detection  \\ 
 \hline 

Edouard A. Hay\cite{ref82} &convolutional neural networks for identification of bacteria: Github repository \url {https://github.com/rplab/Bacterial-Identification.} & Light sheet microscopy image data &over 90\% accuracy  \\ 
\hline 

\end{longtable}

\subsection{Clinical Management and Treatment}

To date, there is no specific treatment for disease caused by the COVID-19 virus. However, many of the symptoms can be treated and therefore, treatment will depend on the patient's clinical condition. Clinical management practices can be improved with practices like classifying patients based on the severity of the disease and providing them with immediate medical care. Due to the multidisciplinary nature of computer vision, it has the potential to support various teams that are currently working on creating vaccination for the disease as well as clinical management.

\subsubsection*{Representative Work, Evaluation and Discussion}

An essential part of the fight against the virus is clinical management, which can be done by identifying patients that are critically ill so that they get immediate medical attention or ventilator support. A disease progression score is recommended to classify different types of infected patients in\cite{ref33}. It is called corona score” and is calculated by measurements of infected areas and the severity of disease from CT images. The corona score measures the progression of patients over time, and it is computed by a volumetric summation of the network-activation maps.\


\begin{figure}[!htb]
        \center{\includegraphics[width=\textwidth]
        {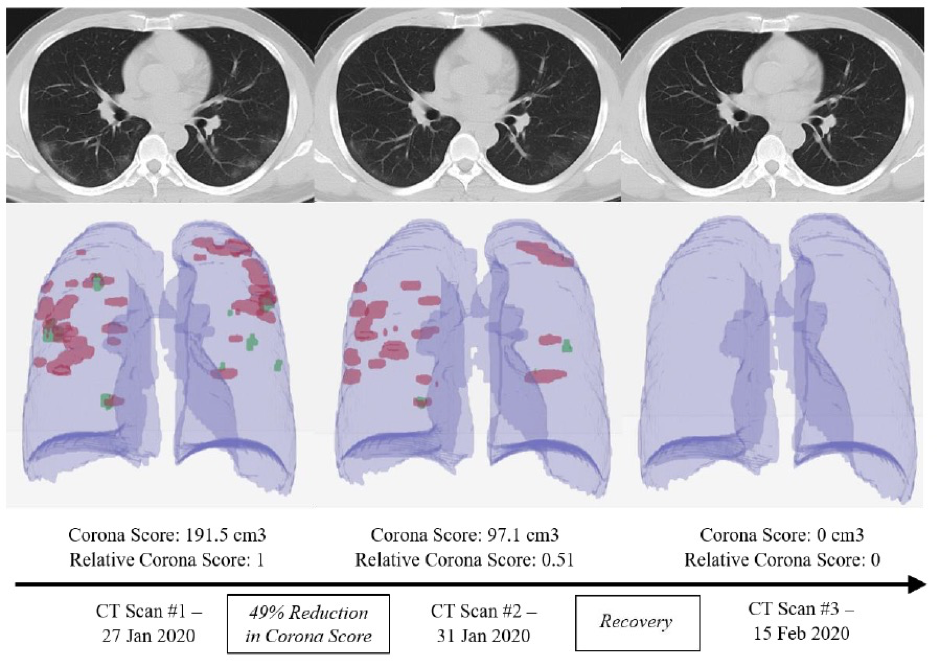}}
        \caption{\label{fig:my-label}: Corona score that is calculated by measurements of infected areas and severity of disease from CT images. It can be used for identifying patients that are critical ill so that they get immediate medical attention. Image adapted from\cite{ref33}.}
      \end{figure}

Graeme MacLaren\cite{ref84} supports that radiological evidence can also be an important tool to distinguish critical ill patients. Yunlu Wang\cite{ref85} used depth camera and deep learning as abnormal respiratory patterns classifier that may contribute to large-scale screening of people infected with the virus accurately and unobtrusively. Respiratory Simulation Model (RSM) is first proposed to fill the gap between the large amount of training data and scarce real-world data. They proposed GRU neural network with bidirectional and attentional mechanisms (BI-AT-GRU) to classify six clinically significant respiratory patterns (Eupnea, Tachypnea, Bradypnea, Biots, Cheyne-Stokes and Central-Apnea) to identify critically ill patients. The proposed model can classify the respiratory patterns with accuracy, precision, recall and F1 of 94.5\%, 94.4\%, 95.1\% and 94.8\%, respectively. Demo videos of this method working in situations of one subject and two subjects can be accessed online (\url{https://doi.org/10.6084/m9.figshare.11493666.v1}).\
  
The CoV spike (S) glycoprotein is a key target for vaccines, therapeutic antibodies, and diagnostics that can guide future decisions. The virus binds to host cells through its trimeric spike glycoprotein. Using biophysical assays, the Daniel Wrapp et al.\cite{ref86} showed that this protein binds at least ten times more tightly than the corresponding spike protein of severe acute respiratory syndrome (SARS)-CoV to their common host cell receptor. These studies provide valuable information to guide the development of medical countermeasures for 2019-nCoV. \

Quantitative structure-activity relationship (QSAR) analysis has perspectives on drug discovery and toxicology\cite{ref87}. It employs structural, quantum chemical and physicochemical features calculated from molecular geometry as explanatory variables predicting physiological activity. Deep feature representation learning can be used for QSAR analysis by incorporating 360° images of molecular conformations into deep learning. Yoshihiro Uesaw\cite{ref88} proposed QSAR (Quantitative structure-activity relationship) analysis using deep learning based on a novel molecular image input technique. Such techniques can be used for drug discovery and can pave the way for vaccine development. \


\begin{longtable}{|L{4cm}|L{6cm}|L{6cm}|} \hline
\caption {\bf : Representative works for infected disease management and treatment}\\
\hline
\bf Study    &  \bf Treatment or Management Methodology      & \bf Implications \\  
\endfirsthead
\hline
\multicolumn{3}{c}%
{\tablename\ \thetable\ -- \textit{Continued from previous page}} \\
\hline
\bf Study&  \bf Treatment or Management Methodology & \bf Implications \\ \hline
\endhead
\hline \multicolumn{3}{r}{\textit{Continued on next page}} \\
\endfoot
\hline
\endlastfoot
\hline

Daniel Wrapp et al.\cite{ref86} & Using biophysical assays, it is shown that this protein binds at least 10 times more tightly than the corresponding spike protein of severe acute respiratory syndrome (SARS)-CoV to their common host cell receptor. & The virus binds to host cells through its trimeric spike glycoprotein. 3.5-angstrom-resolution cryo–electron microscopy structure of the 2019-nCoVS trimer in the prefusion conformation was studied
These studies provide valuable. information to guide the development of medical counter-measures for 2019-nCoV. \\  
\hline

Ophir Gozes et al.\cite{ref33} & Corona score for patient disease progression monitoring and screening & It was basedonthe development of CT images Dataset provide d by ChainZ (www.ChainZ.cn) and Corona score was used to screen critically ill patients. For instance,  Corona score of 191.5 cm3  was observed at the time of Admission and after recovery, it turned 0 means  no opacities \\ 
\hline

Yunlu Wang et. al.\cite{ref85}& Depth camera and deep learnin is used to classify 6 clinically significant respiratory patterns.
GRU neural network with bidirectional and attentional mechanisms (BI-AT-GRU) 
& Abnormal respiratory patterns classifier may contribute to large-scale screening of people infected with COVID-19 in an accurate and unobtrusive manner \\ \hline

Yoshihiro Uesawa et. al.\cite{ref88} & Quantitative structure–activity relationship analysis using deep learning based on a novel molecular image input technique& Drug discovery\\ \hline

\end{longtable}

\begin{figure}[htb]
        \center{\includegraphics[width=\textwidth]
        {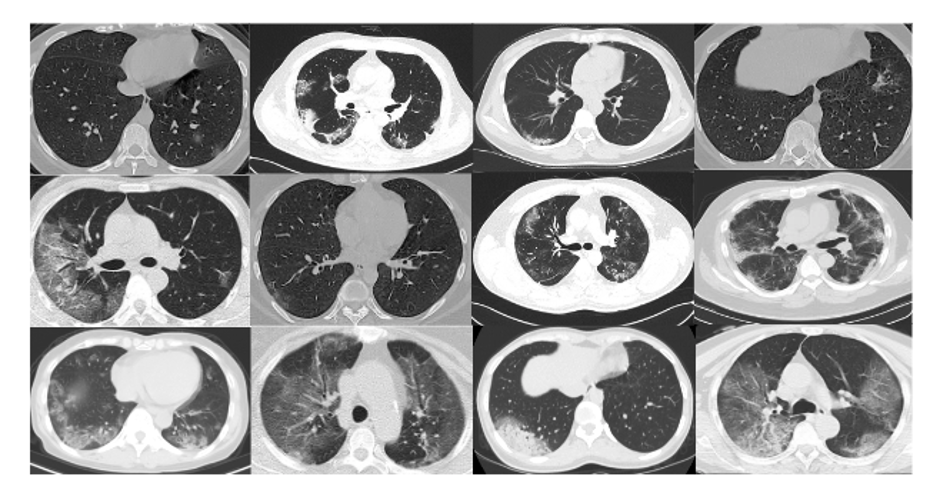}}
        \caption{\label{Figure:my-label} Corona score that is calculated by measurements of infected areas and severity of disease from CT images. It can be used for identifying patients that are critically ill so that they get immediate medical attention. Image adapted from\cite{ref33}.}
      \end{figure}

\section{DATASET AND RESOURCES}

\subsection*{CT images}

\begin{itemize}
    \item COVID-CT-Dataset\cite{ref89}- The University of San Diego has released a data set with 349 CT images containing clinical findings of COVID-19. It claims to be the largest of its kind. To demonstrate its potential, an AI model is trained,  achieving 85\% accuracy. The data set is available at \url{https://github.com/UCSD-AI4H/COVID-CT}. 
    \item An image-based model working with CTs for COVID-19 diagnosis can be found at \url{https://github.com/JordanMicahBennett/SMART-CT-SCAN_BASED-COVID19_VIRUS_DETECTOR/}. 
\end{itemize}{}

\subsection*{CX-ray images}

\begin{itemize}
    \item COVID-19 Radiography database\cite{ref90} - A team of researchers from Qatar University, Doha, and the University of Dhaka, Bangladesh, along with collaborators from Pakistan and Malaysia with medical doctors have created a database of chest X-ray images for COVID-19 positive cases along with Normal and Viral Pneumonia images. In the current release, there are 219 COVID-19 positive images, 1341 normal images and 1345 viral pneumonia images. The authors said that they would continue to update this database as soon as new x-ray images for COVID-19 pneumonia patients. The project can be found at GitHub with MATLAB codes and trained models: \url {https://github.com/tawsifur/COVID-19-Chest-X-ray-Detection}. The research team managed to classify COVID-19, Viral pneumonia and Normal Chest X-ray images with an accuracy of 98.3\%. This scholarly work was submitted to Scientific Reports (Nature), and the manuscript was uploaded to ArXiv. Please make sure to give credit while using the dataset, code and trained models. \

    \item COVID-19 Image Data Collection\cite{ref58}- An initial COVID-19 open image data collection is provided by Joseph Paul Cohen. all images and data are released under the following URL \url{ https://github.com/ieee8023/covid-chestxray-dataset}. \
    
    \item COVIDx Dataset\cite{ref53}- This is the release of the brand-new COVIDx dataset with 16,756 chest radiography images across 13,645 patient cases. The current COVIDx dataset is constructed by the open-source chest radiography datasets at \url{https://github.com/ieee8023/covid-chestxray-dataset} and \url{https://www.kaggle.com/c/rsna-pneumonia-detection-challenge}. It is a combination of data provided by many parties: the Radiological Society of North America (RSNA), others involved in the RSNA Pneumonia Detection Challenge, Dr Joseph Paul Cohen, and the team at MILA, involved in the COVID-19 image data collection project for making data available to the global community. \
    
    \item ChestX-ray8\cite{ref91}- The chest X-ray is one of the most commonly accessible radiological examinations for screening and diagnosis of many lung diseases. A tremendous number of X-ray imaging studies accompanied by radiological reports are accumulated and stored in many modern hospitals' Picture Archiving and Communication Systems (PACS),  available at \url{https://nihcc.app.box.com/v/ChestXray-NIHCC)}.

\end{itemize}{}

\subsection*{Other images}

\begin{itemize}
    \item Masked Face Recognition Datasets\cite{ref71} - Three types of masked face datasets were introduced, including Masked Face Detection Dataset (MFDD), Real-world Masked Face Recognition Dataset (RMFRD) and Simulated Masked Face Recognition Dataset (SMFRD). MFDD dataset can be used to train an accurate masked face detection model, which serves for the subsequent masked face recognition task. RMFRD dataset includes 5,000 pictures of 525 people wearing masks and 90,000 images of the same 525 subjects without masks. To the best of our knowledge, this is currently the world’s largest real-world masked face dataset. SMFRD is a simulated masked face data set covering 500,000 face images of 10,000 subjects. These datasets are available at  \url{https://github.com/X-zhangyang/Real-World-Masked-Face-Dataset.} \
    
    \item Thermal Images  Datasets - There is no dataset of thermals for high fever screening. However, a fully annotated thermal face database and its application for thermal facial expression recognition were proposed by Marcin Kopaczka\cite{ref92}. Information on further ideas of related data that can be figured out by using such systems is available at https://www.flir.com.au/discover/public-safety/thermal-imaging-for-detecting-elevated-body-temperature/.
    
\end{itemize}

\section{CONCLUSION REMARKS}

In this article, we presented an extensive survey of computer vision efforts and methods to combat the COVID-19 pandemic challenge and also gave a brief review of the representative work to date. We divide the described methods into three categories based on their role in disease control: Computed Tomography (CT) scans, X-ray Imagery, and Prevention and Control. We provide detailed summaries of preliminary representative work, including available resources to facilitate further research and development. We hope that, in this first survey on Computer vision methods for COVID-19 control with extensive bibliography content, one can find give valuable insight into this domain and encourage new research. However, this work can be considered only as an early review since many computer vision approaches are being proposed and tested to control COVID-19 pandemic at the current time. We believe that such efforts will be having a far-reaching impact with positive results to periods during the outbreak and post the COVID-19 pandemic.

\bibliographystyle{elsarticle-num} 
\bibliography{bibliography.bib}

\end{document}